\documentclass{acm_proc_article-sp}
\usepackage{local}
\usepackage{stex-logo}
\usepackage{url}
\usepackage{courier}
\usepackage{listings}
\lstMakeShortInline[basicstyle=\tt]|
\usepackage{lstomdoc}
\usepackage{paralist}
\usepackage{amsfonts}
\usepackage[T1]{fontenc}
\usepackage[utf8]{inputenc}
\usepackage[today,eso-foot]{svninfo}
\usepackage{stex-logo}
\newcommand{\latexml}{{\LaTeX}ML\xspace}
\newcommand{\omdoc}{OMDoc\xspace}
\newcommand{\stexide}{\stex\kern-.5em IDE\xspace}
\newcommand{\vmodel}{V-model\xspace}
\newcommand{\sd}{{\small \texttt{SAMS\-Docs}}}
\newcommand{\sams}{\texttt{SAMS}\xspace}

\def\myCitation#1{{\small ``{\emph{#1}}''}}
\def\stexplus{{\stex}{\protect\large{\bf{+}}}\xspace}

\begin{document}
\svnInfo $Id: stex-isem.tex 1265 2010-03-09 10:54:39Z ako $
\svnKeyword $HeadURL: https://svn.kwarc.info/repos/swim/doc/metadata/stex-isem.tex $

%
\conferenceinfo{I-SEMANTICS}{2010, September 1--3, 2010, Graz, Austria}
\CopyrightYear{2010} 
\crdata{978-1-4503-0014-8/10/09}  

\title{\stexplus\ -- a System for Flexible Formalization of Linked Data}
%
%
%
%
%

\numberofauthors{3} 
%
\author{
%
%
\alignauthor
Andrea Kohlhase\\
       \affaddr{German Research Center for Artificial Intelligence (DFKI)}\\
       \affaddr{Enrique-Schmidt-Str. 5}\\
       \affaddr{28359 Bremen, Germany}\\
       \email{Andrea.Kohlhase@dfki.de}
\alignauthor
Michael Kohlhase\\
       \affaddr{Jacobs University Bremen}\\
       \affaddr{P.\,O.\ Box 750561}\\
       \affaddr{28725 Bremen, Germany}\\
       \email{m.kohlhase@jacobs-university.de}
\alignauthor
Christoph Lange\\
       \affaddr{Jacobs University Bremen}\\
       \affaddr{P.\,O.\ Box 750561}\\
       \affaddr{28725 Bremen, Germany}\\
       \email{ch.lange@jacobs-university.de}
}

\maketitle
\begin{abstract}
  We present the \stexplus system, a user-driven advancement of {\stex} --- a semantic
  extension of {\LaTeX} that allows for producing high-quality PDF documents for
  (proof)reading and printing, as well as semantic XML/{\omdoc} documents for the Web or
  further processing.  Originally {\stex} had been created as an invasive, semantic
  frontend for authoring XML documents.  Here, we used \stex in a Software Engineering
  case study as a formalization tool. In order to deal with modular pre-semantic
  vocabularies and relations, we upgraded it to \stexplus in a participatory design
  process.  We present a tool chain that starts with an \stexplus editor and ultimately
  serves the generated documents as XHTML+RDFa Linked Data via an \omdoc-enabled,
  versioned XML database.  In the final output, all structural annotations are preserved in order to enable semantic information retrieval services.
\end{abstract}




\section{Introduction}

An important issue in the Semantic Web community was and still is the ``Authoring
Problem'': How can we convince people not only to use semantic technologies, but also prepare
them for creating semantic documents (in a broad sense)? Here, we were interested in
formalizing a collection of {\LaTeX} documents into a set of files in the {\omdoc} format,
an XML vocabulary specialized for managing mathematical information, and further on to
Linked Data for interactive browsing and querying on the Semantic Web.

\begin{figure}[h]
\centering
  \includegraphics[width=.4\textwidth]{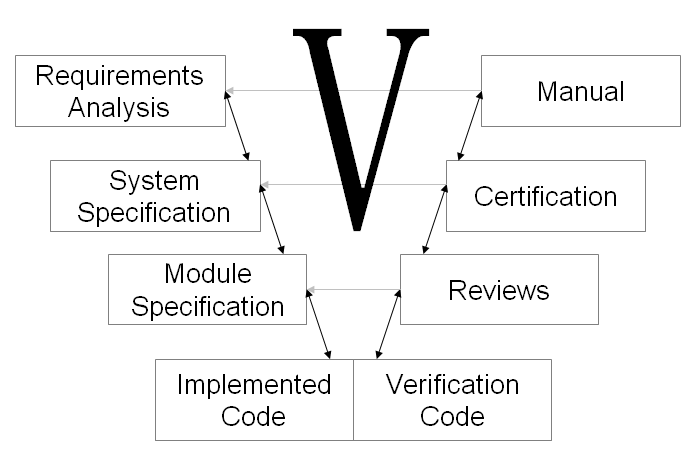}    
  \caption{A Document View on the V-Model}\label{fig:V-model}
\end{figure}

Concretely, the object of our study was the collection of documents created in the course
of the 3-year project ``Si\-che\-rungs\-kom\-po\-nen\-te f\"ur Autonome Mobile Systeme
({\bf{\sams}})'' at the German Research Center for Artificial Intelligence (DFKI). {\sams}
built a software safety component for autonomous mobile service robots developed and certified it
as SIL-3 standard compliant (see~\cite{sams:SafeCert08}).  Certification required the
software development to follow the {\vmodel} (figure~\ref{fig:V-model}) and to be based on
a verification of certain safety properties in the proof checker
Isabelle~\cite{Nipkow-Paulson-Wenzel:2002}. The {\vmodel} mandates e.\,g.\ that relevant
document fragments get justified and linked to corresponding fragments in other members of
the document collection in an iterative refinement process (the arms of the `V' from the
upper left over the bottom to the upper right and in-between in figure~\ref{fig:V-model}).

System development with respect to this regime results in a highly interconnected
collection of design documents, certification documents, code, formal specifications, and
formal proofs.  This collection of documents ``{\textbf{\sd}}''~\cite{SAMSDocs} make up
the basis of a case study in the context of the FormalSafe project~\cite{URL:FormalSafe}
at DFKI Bremen, where they serve as a basis for research on machine-supported change
management, information retrieval, and document interaction.  In this paper, we report on
the formalization project of the collection of {\LaTeX} documents in {\sd} (that we will
without further ado also abbreviate with {\sd}).

Not surprisingly, the interplay between the fields Semantic Web and Human-Computer
Interaction played an important role as the ``Authoring Problem'' of the
first is often tackled via methods of the second. One such approach is that of
{\myCitation{invasive technology}}~\cite{Kohlhase:ophcie05} with the basic idea that from
a user's perspective, semantic authoring and general editing are the same, so why not
offer semantic functionalities as an extension of well-known editing systems, thereby
`invading' the existent ones. We started with {\LaTeX} not only because a good portion of
our case study was written in it, but also as {\LaTeX} constitutes the state-of-the art
authoring solution for many scientific/technical/mathematical document
collections. Despite its text-based nature it is widely considered the most efficient tool for the
task. Therefore, we used the invasive OMDoc frontend for {\LaTeX} documents called
{\textbf{\stex}}~\cite{Kohlhase:ulsmf08}. In the formalization process its conceptual
usability weaknesses (for the task) were identified and within a participatory design process
it evolved into the invasive formalization tool {\stexplus}.

In section~\ref{sec:stex}, we will present the \stex system, especially its realization of
Linked Data creation. Then we describe in section~\ref{sec:samsdocs} the formalization
process of {\sd} with \stex, our challenges, and our (pre-)solutions. In
section~\ref{sec:stex-new} we report the enhancements of {\stex} realized in and for the
case study to {\stexplus}. Having {\stexplus} documents with Linked Data and ontological
markup, we describe (potential) services and their implementation design in
section~\ref{sec:services}. Section~\ref{sec:related} summarizes related work, and section~\ref{sec:conclusion} concludes the paper.

\section{\protect\stex: Obj.-Oriented {\LaTeX} Markup}\label{sec:stex}

{\stex}~\cite{Kohlhase:ulsmf08,sTex:web} is an extension of the {\LaTeX} language that is geared towards
marking up the semantic structure underlying a document.  The main concept in \stex is
that of a ``{\textbf{semantic macro}}'', i.\,e., a {\TeX} command sequence $\mathcal{S}$
that represents a meaningful (mathematical) concept $\mathcal{C}$: the {\TeX} formatter
will expand $\mathcal{S}$ to the presentation of $\mathcal{C}$. For instance, the command
sequence |\positiveReals| (from listing~\ref{lst:stex-ex}) is a semantic macro that
represents a mathematical symbol --- the set $\mathbb{R}^+$ of positive real
numbers. While the use of semantic macros is generally considered a good markup practice
for scientific documents (e.\,g., because they allow to adapt notation by macro redefinition
and thus increase reusability), regular {\TeX/\LaTeX} does not offer any infrastructural
support for this. \stex does just this by adopting a semantic, `object-oriented' approach
to semantic macros by grouping them into ``modules'', which are linked by an ``imports''
relation.  To get a better intuition, consider
\begin{lstlisting}[label=lst:stex-ex,caption=An \protect\stex Module for
  Real Numbers,escapechar=|,language=sTeX]
\begin{module}[id=reals]
  \importmodule[../background/sets]{sets}
  \symdef{Reals}{\mathbb{R}}
  \symdef{greater}[2]{#1>#2}
  \symdef{positiveReals}{\Reals^+}
  \begin{definition}[id=posreals.def,
   title=Positive Real Numbers]
    $\defeq\positiveReals
             {\setst{\inset{x}\Reals}{\greater{x}0}}$
  \end{definition}
  |\ldots|
\end{module}
\end{lstlisting}
which would be formatted to

\vspace{1ex}\hrule
\begin{small}
\noindent\textbf{Definition} 2.1 (Positive Real Numbers): $\mathbb{R}^+:=\{x\in\mathbb{R}\mid x>0\}$
\end{small}\hrule\vspace{1ex}

Here, {\stex}'s |\symdef| macro -- invasive by to its deliberate resemblance of (La){\TeX}'s |\def| and |\newcommand| -- generates a respective semantic macro, for
instance the |\positiveReals| with representation  $\mathbb{R}^+$.
Note the symbol inheritance scheme of {\stex}: The markup in the module |reals| has access to semantic macros
|\setst| (``set such that'') and |\inset| (set membership) from the
module |sets| that was imported by the document |\importmodule| directive from
the \url{../background/sets.tex}. Furthermore, it has access to the |\defeq|
(definitional equality) that was in turn imported by the module |sets|. 

From this example we can already see an organizational advantage of \stex over {\LaTeX}:
we can define the (semantic) macros close to where the corresponding concepts are defined,
and we can (recursively) import mathematical modules. But the main advantage of markup in
\stex is that it can be transformed to XML via the {\latexml}
system~\cite{Miller:latexml}: Listing~\ref{lst:omdoc-ex} shows the
{\omdoc}~\cite{Kohlhase:OMDoc1.2} representation generated from the {\stex} sources in
listing~\ref{lst:stex-ex}. {\bf{\omdoc}} is a semantics-oriented representation format for
mathematical knowledge that extends the formula markup formats
OpenMath~\cite{BusCapCar:2oms04} and MathML~\cite{CarlisleEd:MathML3} to a document markup
format.

\lstset{language=[1.6]OMDoc}
\begin{lstlisting}[label=lst:omdoc-ex,mathescape,escapeinside={\{\}},caption={An XML Version of Listing~\ref{lst:stex-ex}}]
<theory xml:id="reals">
 <imports from="../background/sets.{omdoc}#sets"/>
 <symbol xml:id="Reals"/>
 <notation>
   <prototype><OMS cd="reals" name="Reals"/></prototype>
    <rendering><m:mo>$\mathbb{R}$</m:mo></rendering>
 </notation>
  <symbol xml:id="greater"/><notation>$\ldots$</notation>
  <symbol xml:id="positiveReals"/><notation>$\ldots$</notation>
  <definition xml:id="posreals.def" for="positiveReals">
    <meta property="dc:title">Positive Real Numbers</meta>
    <OMOBJ>
      <OMA>
        <OMS cd="mathtalk" name="defeq"/>
        <OMS cd="reals" name="positiveReals"/>
        <OMA> 
           <OMS cd="sets" name="setst"/>
           <OMA>
            <OMS cd="sets" name="inset"/>
            <OMV name="x"/>
            <OMS cd="reals" name="reals"/>
           </OMA>
           <OMA>
             <OMS cd="reals" name="greater"/>
             <OMV name="x"/>
             <OMI>0</OMI>
          </OMA>
        </OMA>
      </OMA>
    </OMOBJ>
 </definition>
   $\ldots$
</theory>
\end{lstlisting}
One thing that jumps out from the XML in this listing is that it incorporates all the
information from the \stex markup that was invisible in the PDF produced by formatting it
with {\TeX}.  

\begin{figure}[ht]\centering
  \includegraphics[width=\linewidth]{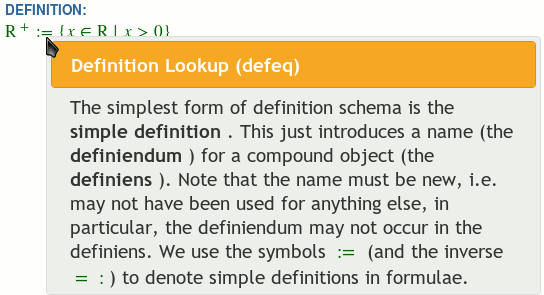}    
\caption{Listing~\ref{lst:stex-ex} as Dynamic XHTML+MathML}\label{fig:xhtml+mathml-ex}
\end{figure}

{\omdoc} itself has been used as a storage and exchange format for automated theorem
provers, software verification systems, e-learning software, and other
applications~\cite[chapter~26]{Kohlhase:OMDoc1.2}, but due to its focus on semantic
structures, it is not intended to be consumed by human readers.
The Java-based JOMDoc~\cite{JOMDoc:web} library uses the |notation| elements to
generate human-readable XHTML+MathML from OMDoc.  Figure~\ref{fig:xhtml+mathml-ex} shows
the result of rendering the document from listing~\ref{lst:omdoc-ex} in a MathML-aware
browser.  In contrast to the PDF output we can directly create from \stex, XHTML+MathML
allows for interactivity. In particular, our JOBAD JavaScript framework enables modular interactive services in rendered XHTML+MathML documents~\cite{GLR:WebSvcActMathDoc09}.
These services utilize the semantic structures of mathematical
formulae.  In our rendered documents, each formula in
human-readable Presentation MathML carries the original semantic OpenMath representation
of the formula, as shown in listing~\ref{lst:omdoc-ex}, as a hidden annotation.

Client-side JOBAD services, which exclusively rely on annotations given inside a document, have already been implemented for folding and unfolding subterms of
formulae and for controlling the display of redundant brackets in complex formulae.  The symbol definition lookup
service, shown in figure~\ref{fig:xhtml+mathml-ex}, interacts with a server backend: It
traverses the links to |symbol| and their corresponding |definition|
elements that are established by the |OMS| elements in OpenMath – for example,
\lstinline[mathescape,language=OpenMath,basicstyle=\tt]{<OMS cd="sets"$\mathtt{~}$name="inset"/>} encodes
the URI \url{../background/sets.omdoc#inset} – and retrieves the document at that URI as
XHTML+Math\-ML.\footnote{This is the MathML way of representing Linked Data.  In
  section~\ref{sec:services}, we describe how we have now extended this feature to cover RDFa
  Linked Data.}  JOBAD's ability to integrate an arbitrary number of services, which can
talk to different server backends and which are enabled depending on the context, i.\,e., the
semantic structure of the part of a mathematical formula that the user has selected, turns
our rendered mathematical documents into powerful mashups~\cite{KGLZ:JOBADabstract09}.  On
any symbol, for example, definition lookup is enabled.  On any expression where a number
is multiplied with a special symbol representing a unit of measurement, a unit conversion client
that talks to a remote unit conversion web service is enabled.  The JOBAD architecture has
been designed without depending on a particular backend; for most of our services we are
using the extensible XML-aware database TNTBase~\cite{ZhoKoh:tvsx09,ZhoKohRab:tntbasef10,DKLRZ:PubMathLectNotLinkedData10}, which has special support for OMDoc and
integrates the JOMDoc rendering library.

\section{Formalization with \protect\sTeX towards \protect\stexplus}\label{sec:samsdocs}
In this section we describe the process of formalizing the {\sd} collection of {\LaTeX}
documents created in the course of the {\sams} project with the \stex system. We use the
user's perspective to point to the requirements for {\stexplus} that evolved in this process.

As we all know all too well: Formalizing is never easily done. In our project we had the
additional challenge of doing it without corruption of the PDF layout that was produced
with {\LaTeX}. Here, \stex fits well, as it generates PDF and transforms to XML.
\begin{figure*}[ht]
\centering
  \includegraphics[width=0.8\textwidth]{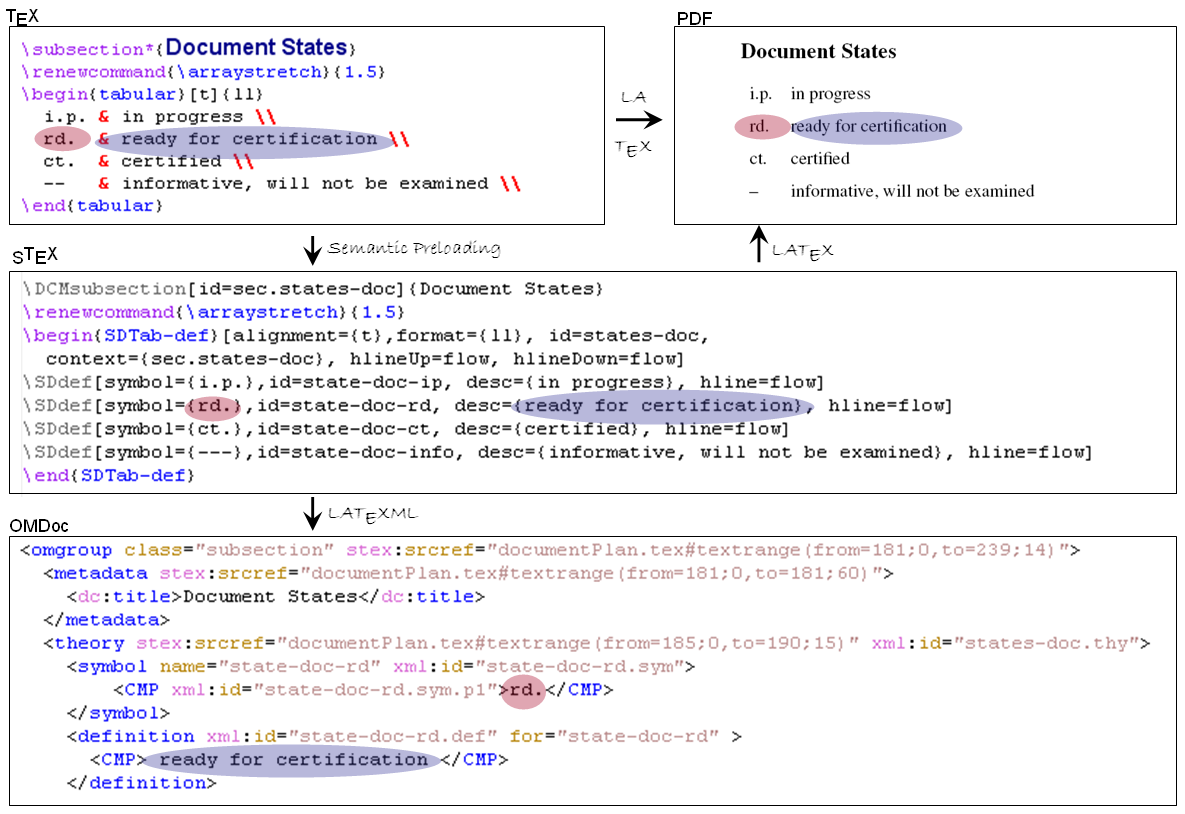}    
  \caption{The Formalization Workflow via \protect\stex: Definition Table of ``document state''}\label{fig:workflow}
\end{figure*}
In figure~\ref{fig:workflow} we can see the general course of action: 
\begin{compactenum}[\em i)]
\item we identified document fragments (``{\textbf{objects}}'') that constitute a coherent,
meaningful unit like the state of a document ``rd.\@'' or its description ``ready for certification'', then 
\item we translated it into the \stex format, realizing for example that ``rd.\@'' is a
  recurring symbol and ``ready for certification'' its definition (therefore designing the {\sd} macro ``|SDdef|''), and finally
\item we polished these macros in the \stex specific sty-files so that the PDF layout
  remained as before and the generated XML represented the intended logical structure, for
  instance the use of the {\omdoc} XML elements |symbol| and |definition|.
\end{compactenum}
Note that definitions are common objects in mathematical documents, therefore \stex
naturally provides a {|definition|} environment. So why didn't we use that?  Because the
document model of {\omdoc}, which we obtain by transforming {\stex} using {\latexml}, does not
allow definitions in tables, as the former are stand-alone objects from an ontological
perspective. If one {\emph{authors}} a formal document, this view is taken, so no problem
arises, but if one {\emph{formalizes}} an existing document, layout and cognitive
side-conditions have to be taken into account. We therefore realized that we could not
simply add basic \stex markup to the {\LaTeX} source yielding formal objects, we rather
needed to add pre-formal markup in the formalization process (we speak of
{\textbf{(semantic) preloading}}).

Whenever project-wide (semantic) layout schemes were discovered, that were frequently used, we
extended the macro set of {\stex} suitably (enabling preloading ``{\emph{project
    structures}}''~\cite{KohKohLan:difcsmse10}, i.\,e.\ project-induced ones which is quite
different from ``{\emph{document [layout] structures}}''~[ibid.], e.\,g.\ by
subsections that is supported by {\stex} core features, see {|DCMsubsection|} in
figure~\ref{fig:workflow}). The table layout for example was often used for lists of symbol
definitions. So we created the {|SDTab-def|} environment which can host as many {|SDdef|}
commands as wanted (see fig.~\ref{fig:workflow}). This increased the efficiency of the
formalizing process tremendously.

Another difference between authoring and semantic preloading consisted in the {\emph{order
    of the formalization steps}}. While the order of the first typically consists of
``{\textbf{chunking}}'' (i.\,e., building up structure e.\,g.\ by setting up theories),
``{\textbf{spotting}}'' (i.\,e., coining objects), and ``{\textbf{relating}}'' (i.\,e.,
making relationships between objects or structures explicit), the order of the second is made
up of spotting, then relating {\emph{or}} chunking. The last two were done simultaneously,
because \stex offers a very handy inheritance scheme for symbol macros --- as long as
the chunks are in order, which could be sensibly done for some but not for all at this
stage in the formalization process. Generally, many `guiding' services of \stex, that
\stex considered to be features, turned out to be too rigid.

As a consequence we heavily used very light annotations at the beginning: It was sufficient to
identify a certain document fragment and to mark it with a referencable ID like
``state-doc-rd''. Shortly afterwards, we realized that some more basic markup was necessary, since
we wanted to formalize our knowledge of types/categories of these objects and their conceptual
belonging. For this we developed a set of ``{\textbf{ad-hoc semantification macros}}'' with named
attributes like |SDobject[id]|, |SDmore[id,cat,for]|,\\
|SDisa[id,cat,for,follows,theory,imports,tab]|, or |SDreferences[id,file,refid]|\footnote{We use
  subsets of a general attributes set for all of our {\stex} extensions to lower the learning curve
  for the use of the markup macros.}.  The `more' functionality provided by |SDmore| was required
due to logically contiguous objects that were interspersed in a document. With this set we preloaded
``{\emph{object structures}}''~[ibid.], i.e. object-induced ones. Note that the ad-hoc semantification macros enabled the formalizer to
develop her own metadata vocabulary.

As soon as the document boundaries went down, we realized that an object had many
occurrences in several of the documents in the {\sd} collection. For example, first an
object was introduced as a high-level concept in the contract, then it was specified in
another document, refined in a detailed specification, implemented in the code, reviewed
at some stage, and so on until it was finally described in the manual. Thus, we had to
preload ``{\emph{collection structures}}''~[ibid.] as well, which consisted in the development process
model, the {\vmodel} as seen in figure~\ref{fig:V-model}. Here, we built our personal
{\bf{{\vmodel} macros}}, e.\,g.\ {|SemVMrefines|}, {|SemVMimplements|}, or
{|SemVMdescribesUse|}.

\begin{figure}
\centering
  \includegraphics[width=\linewidth]{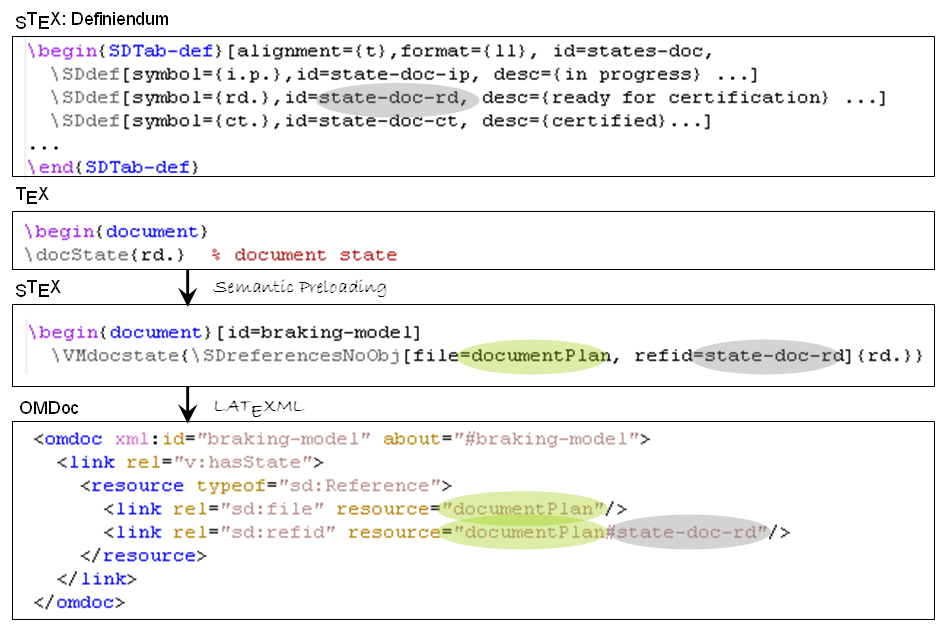}
  \caption{Referencing a ``document state''}\label{fig:workflow-ref}
\end{figure}

Additionally, we created an {\stex} extension especially suited for preloading
``{\emph{organizational structures}}''~[ibid.]. This is considered different from project structures
as organizational markup is very probable to be reusable for other projects with the same
organizational structures. For example, {\sams} used a document version management as well
as a document review history, so that environments |VMchangelist|, |VMcertification| with corresponding list entry macros
|VMchange|, |VMcertified| were built. Another example is the processing state of a
document, which can be marked up easily by using the |VMdocstate| macro as seen in
figure~\ref{fig:workflow-ref}.

We noted that the necessary formalization depth of some documents was naturally deeper
than others. For example, it didn't seem sensible to formalize the contract too
much, as it was created as a high-level communication document, whereas the detailed specification
needed a lot of formalization. The manual had an interesting mixed state of formality and
informality, as it was again geared towards communication, but it needed to be very
precise.  In conclusion we note that the mathematical content of the documents (i.\,e., the
mathematical objects and their relations) was only one of the knowledge sources that
needed to be formalized and marked up.  In the course of the formalization it has become
apparent that the knowledge in such complex collections is \emph{multi-dimensional}
(cf.~\cite{KohKohLan:difcsmse10} for an in-depth analysis).  Thus, the requirements for extending {\stex} to {\stexplus} were \begin{inparaenum}[(i)]\item to generate XML output that preserves the semantics annotated in the preloading phase, \item and to take into account the multi-dimensionality of our ad-hoc semantification macros in a way that technically enables browsing and querying\end{inparaenum}.  These requirements were satisfied by enabling the generation of RDFa from our annotations and making them accessible to Linked Data services, as we will describe in the following sections.

\section{\protect\stexplus: a Metadata-Extension of \protect\stex}\label{sec:stex-new}

All the arrows in figure~\ref{fig:V-model} are
examples of relations between document fragments in the {\sd} corpus that needed to be
made explicit in addition to the mathematical relations that {\stex} had originally supported; the revision histories of documents and the social networks of their authors constitute further dimensions of knowledge.  For situations like these, we had incorporated RDFa~\cite{w3c:rdfa-primer} as a flexible metadata
framework into the {\omdoc} format~\cite{LK:MathOntoAuthDoc09}.  In the course of this case
study, the RDFa integration was revised and extended and will become part of the upcoming
{\omdoc} version 1.3~\cite{Kohlhase:OMDoc1.3}. The main idea for this integration is to
realize that any concrete document markup format can only treat a certain set of objects
and their relations via its respective native markup infrastructure.  All other objects
and relations can be added via RDFa annotations to the host
language -- assuming the latter is XML-based.

It is crucial to realize that, for machine support,
the metadata objects and relations are given a machine-processable meaning via suitable
ontologies.  Moreover, ontologies are just special cases of (mathematical) theories, which
import appropriate theories for the logical background, e.\,g.\ description logic, and
whose symbols are the entities (class, properties, individuals) of ontologies.  Thus,
\stex and {\omdoc} can play a dual role for Linked Data in documents with mathematical
content. They can be used as markup formats for the documents and at the same time as the
markup formats for the ontologies. We have explored this correspondence for {\omdoc} in
previous work and implemented a translation between {\omdoc} and
OWL~\cite{LK:MathOntoAuthDoc09,Lange:PhD}.

To understand our contribution, note that we can view {\LaTeX} and {\stex} as frameworks
for defining domain-specific vocabularies in classes and packages; {\LaTeX} is used for
layout aspects, and \stex can additionally handle the semantic aspects of the
vocabularies. \stex uses this approach to define special markup e.\,g.\ for definitions (see
lines 10 to 31 in listing~\ref{lst:omdoc-ex}). Note that to define \stex markup
functionality like the |definition| environment, we have to provide a {\LaTeX}
environment {\texttt{definition}} (so that the formatting via {\LaTeX} works) and a {\latexml}
binding (to specify the XML transformation for the |definition| environment). As
the {\omdoc} vocabulary is finite and fixed, \stex can (and does) supply special {\LaTeX}
macros and environments and their {\latexml} bindings. But the situation is different for
the flexible, RDFa-based metadata extension in \omdoc1.3 we mentioned above, with a potentially infinite supply of vocabularies. At the start
of the {\sd} preloading effort, \stex already supported a common subset of metadata
vocabularies. For instance the Dublin Core |title| metadata element in line 11 of
listing~\ref{lst:omdoc-ex} is the transformation result of using the
KeyVal~\cite{Carlisle:tkp99} pair \lstinline[mathescape,basicstyle=\tt]|title=$\ldots$| in the optional
argument of the |definition| environment.

\begin{figure}
\centering
  \includegraphics[width=.8\linewidth]{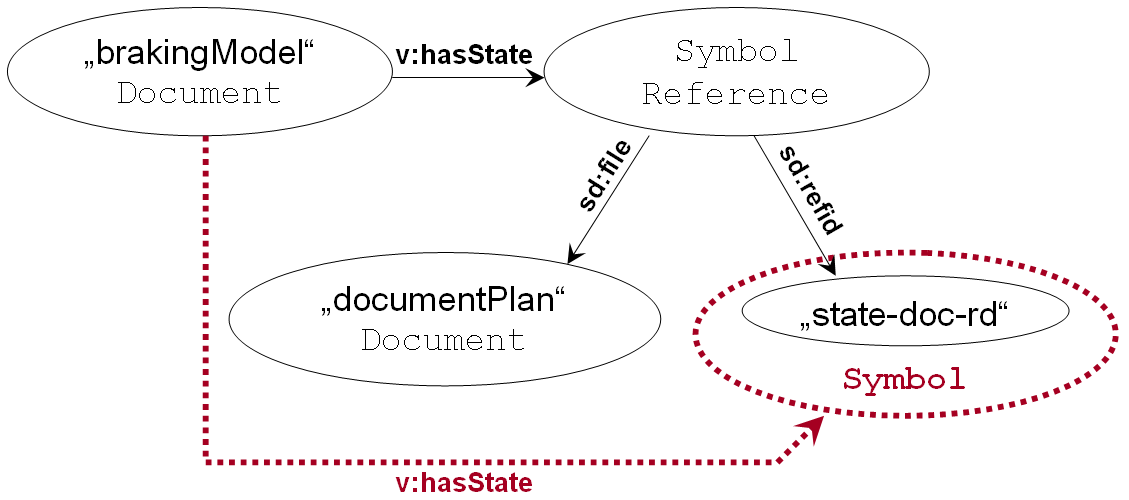}
  \caption{RDF View on a ``doc.\ state'' Assignment}\label{fig:workflow-symbolref}
\end{figure}

For the {\sd} case study we started in the same way by adding a package with {\latexml}
bindings to \stex. The |\VMdocstate| macro shown in the ``{\stex}'' box of
figure~\ref{fig:workflow-ref} allowed us to annotate a document with its processing state.
This is transformed to an RDFa-annotated |omdoc| root element, as shown in the
``{\omdoc}'' box underneath and in the black, solid parts of the RDF graph in
figure~\ref{fig:workflow-symbolref}.  We can already see that the \stex extension for {\sd}
exactly consists in a domain-specific metadata vocabulary extension, and that using the
custom vocabulary hides markup complexity from the author. Again, {\sd} only needed a finite
vocabulary extension, so this approach was feasible, but of restricted applicability,
since developing the {\sd} package for \stex required insights into \stex internals and
\latexml bindings. Thus this extension approach lacks the flexible user-extensibility that would be needed to scale up further.

To enable user-extensibility, we add a new declaration form |\keydef| to the core \stex
functionality (yielding \stexplus) --- like |\symdef| in that it is inherited via the module imports relation,
only that it defines a KeyVal key instead of a semantic macro. To understand its
application, we rationally reconstruct the |v:hasState| relation from the example in
the \omdoc box of figure~\ref{fig:workflow-ref}. To do this, we use \stex to create a
metadata vocabulary for document states: we create a |certification| module, which
defines the |hasState| metadata relation and adds it to the KeyVal keys of the
|document| environment. The |metalanguage| macro is a variant of |importmodule| that imports
the meta language, i.\,e., the language in which the meaning of the new symbols is
expressed; here we use OWL.

\begin{lstlisting}[language=sTeX,caption=A Metadata Ontology for Certification,label=lst:certification]
\begin{module}[id=certification]
 \metalanguage[../background/owl]{owl}
 \keydef{document}{hasState}
 \symdef{state-doc-rd}[1]{rd. #1}
 \symdef{tuev}{\text{T\"UV}}
 \begin{definition}[for=hasState]
  A document {\definiendum[hasState]{has state}} $x$, iff 
  the project manager decrees it so. 
 \end{definition}
 \begin{definition}[for=state-doc-rd]
  A document has state \definiendum[state-doc-rd]{rd. $x$}, 
  iff it has been submitted to $x$ for certification.
 \end{definition}
 \begin{definition}[for=tuev,hasState=$\statedocrd\tuev$]
   The $\tuev$ (Technischer \"Uberwachungsverein) is a 
   well-known certification agency in Germany.
 \end{definition}
\end{module}
\end{lstlisting}

In this paper, we focus on using {\stexplus} as a language for defining lightweight vocabularies.  Note, however, that ``heavyweight'' formal semantics can be added to vocabulary terms in the same way as has been shown for mathematical symbols in listing~\ref{lst:stex-ex}.  Similarly as the ``real numbers'' module relies on an {\stex} module that introduces set theory, the certification ontology relies on an {\stex} module that introduces the OWL language.  Such an OWL ontology that has been written in {\stexplus} can be translated to one of the widely supported serializations of OWL via two paths:  \begin{inparaenum}[(i)]\item In the original workflow, the {\stexplus} source is translated to {\omdoc}.  Thanks to their modularity and literal programming capabilities, the {\stexplus} or {\omdoc} representation allows for an expressive documentation of OWL ontologies.  But, as {\omdoc} is not universally understood on the Semantic Web, we have implemented a translation of OWL ontologies encoded and documented in {\omdoc} to the standard RDF/XML representation~\cite{LK:MathOntoAuthDoc09}.  \item Alternatively to this previously existing translation via {\omdoc} as an intermediate representation, we are working on a direct {\stexplus} to OWL transformation.  Simply using our experimental \url{owl2onto} class~\cite{Kohlhase:owl2onto*:web}
instead of the \url{omdoc} class from \stex in the {\LaTeX} preamble will cause {\latexml}
to generate OWL -- here in the direct OWL XML serialization -- instead of \omdoc from a subset of the {\stexplus} markup.\end{inparaenum}

\begin{lstlisting}[language=sTeX,caption=Annotating a Document with Certification Metadata,label=lst:use-cert]
\importmodule[../ontologies/cert]{certification}
\begin{document}[hasState=$\statedocrd{\tuev}$]
...
\end{document}
\end{lstlisting}

Let us now see how to \emph{use} a vocabulary: If we import the |certification| metadata
module, we can write to generate RDFa annotations that correspond to the (red) dotted
arrow in figure~\ref{fig:workflow-symbolref}. Note that in the state of formalization
shown in figure~\ref{fig:workflow-ref}, the {\sd}-specific RDF vocabulary still has a
pre-semantic structure. With the \stexplus we can express that the processing state is
actually intended to be a symbol in a metadata theory, not just some semantic object in
some file. In listing~\ref{lst:certification} we use the |\symdef| directive to generate
the symbol |state-doc-rd| and |\keydef| to generate a metadata relation |hasState| that is
expressed by a key of the same name, which is added to the |document| environment.  When
processed by {\latexml}, |\keydef| takes care of generating correct URIs for the metadata
relations and their target resources, resulting in an RDFa output syntactically similar to
figure~\ref{fig:workflow-ref}.  In conclusion, we note that \stexplus allows us to
rationally recreate the effect we previously achieved with the custom |\VMdocstate| and
|\SD| |referencesNoObj| macros.  Note that we did not have to extend the {\latexml}
bindings at all for this extension.  Thus, {\stexplus} gives us a generic {\TeX}$\to$RDFa translation, which works for arbitrary vocabularies.\footnote{Our experimental \url{rdfameta} package~\cite{Kohlhase:rdfmeta*:web} extends this to arbitrary {\LaTeX} documents:  It redefines common {\LaTeX} commands (e.\,g.\ the
sectioning macros) so that they include optional KeyVal arguments that can be extended by
\texttt{\textbackslash keydef} commands. With this metadata extension, we can add RDFa metadata to \emph{any} existing {\LaTeX}.}

\section{\protect\stexplus Documents as Linked Data}\label{sec:services}

The translation of classical \stex to \omdoc and further to XHTML+MathML (see
section~\ref{sec:stex}), which results in a Linked Data like markup for mathematical symbols, enables interactive services in mathematical formulae.  Now
that \stexplus supports formalization with arbitrary metadata (cf.\
section~\ref{sec:stex-new}), it should additionally be possible to utilize \emph{these} metadata
for services.  Both types of annotation complement each other:  A practical {\stexplus} document, like many of the {\sd}, would combine elements from listing~\ref{lst:use-cert} with those from listing~\ref{lst:stex-ex} and consequently rely on services for both types of semantic structures.

The JOBAD service architecture (see section~\ref{sec:stex}) gives uniform
access to common queries in the document browsing user interface. In the {\sd} scenario
this might be a query for all persons who have worked on the current document. This can
directly be answered from the metadata of the revision log. Another typical query would
consist in asking for all parts of a specification that have to be re-certified. Answering
this query involves revision logs (for finding documents that have changed since the last
certification), collection structures (\vmodel dependencies of changed parts), and mathematical structures (logical dependencies).  In~\cite{KohKohLan:difcsmse10} we have
elaborated on such {\sd} queries from the point of view of their stakeholders (like
engineers, project managers, certifiers), particularly exploring the multi-dimensional\-ity
of the formal structures.  For example, a project manager may find a substitute for an employee $E$, who has implemented a specification, by tracing back a link from the documentation of the implementation to the specification document and finding out, from the metadata of that document, who has recently been working on it.  Here, we will summarize the extensions made to our system
architecture to enable these services.  

As a first step, we made the JOMDoc renderer preserve the RDFa metadata from the OMDoc documents,
now generating XHTML+MathML+RDFa. Additionally, the mathematical structures (those that are above the formula level) had to be preserved in
the rendered output. Even though OMDoc uses native non-RDFa markup for these structures, we can also represent these in RDF, exploiting
the OMDoc ontology
(see~\cite{OMDocDocOnto:web,DKLRZ:PubMathLectNotLinkedData10} for more information).  Existing JOBAD
services recognized mathematical formulae in XHTML presentations of \omdoc documents by their
semantic structure (e.\,g.\ whether they use previously defined symbols or units of measurement). Similarly,
new services can now recognize from the RDFa annotations whether a chunk of an XHTML document is, e.\,g., an implementation of a
specification fragment, and by which user requirement that is induced.  Compared to the previously existing definition lookup service, the principle of retrieving content from a target URI and displaying it in a popup
remained the same -- the URIs are just provided by different annotations.  

Secondly, we have extended the folding of subterms of mathematical formulae to higher-level
structures, such as requirements or steps of structured proofs.  We have
implemented this using the rdfQuery JavaScript library~\cite{rdfQuery:web}, which parses all RDFa
annotations of a document into a local triple store that can be queried using SPARQL-like JavaScript
functions.  On the server side, we have extended TNTBase~\cite{ZhoKoh:tvsx09}, our versioned
database backend and web server/application framework to accept commits of \stexplus documents,
automatically convert them to OMDoc, and then serve OMDoc, XHTML+MathML+RDFa, and, optionally,
RDF/XML, according to the Linked Data best practices~\cite{LinkedDataGuidesTutorials:web}.

Even the pre-semantic annotations like the ones shown in figure~\ref{fig:workflow-ref} 
afford interactive services: A generic reference can already be utilized for lookup and
navigation. Providing additional information in the instance document or in the
ontology (e.\,g.\ the knowledge about the target of a reference being a symbol or a
processing state) allows for making the service user interface more specific and
enables the display of more relevant related information. For the generic pre-semantic ``references'' relation, the list of all semantic objects that it relates to each other would be too large for being usable, as there is no obvious way of ranking or filtering the link targets.  But once more specific link types are used, such as the ``has state'' link, that information can be used to display a list of documents grouped by state.

Queries across documents cannot be answered using the above-mentioned rdfQuery: client
side queries require a combination of querying a local triple store and crawling links. In
our setup, we have experimented with SQUIN~\cite{SQUIN:web}, a frontend to the Semantic
Web Client library~\cite{SemWebClientLib:web}, which gives access to Linked Data via a
simple HTTP frontend at very low integration costs: If the server provides
standard-compliant Linked Data, then the client simply has to access the URL of the SQUIN
server, providing a SPARQL query as a parameter.  An alternative would have been AJAR
library, a part of the Tabulator Linked Data browser~\cite{tbl:tabulator}, which
implements the same functionality in JavaScript. In our test setup, SQUIN acted as a proxy
between the client-side JavaScript code and our Linked Data. While a Linked Data crawler
is most flexible when data are distributed across many servers (e.\,g.\ when an OMDoc
document links to DBpedia), its query answering capabilities are only as good as the
Linked Data being served. For example, if the RDF(a) does not contain back-links (like
links from a mathematical theory to the theories it imports {\emph{and}} to the theories
by which it is imported), then an AJAR- or SQUIN-powered client cannot query links in both
directions.  Moreover, the performance of such a solution is limited, as it requires
memory for the local triple store as well processor time for query answering on the client
side. Therefore, in the {\sd} setting, where the queries are currently limited to a document
collection on a single server, the best solution is storing the triples on that same
server, and making them accessible via a standard query interface. Concretely, we make a
SPARQL endpoint powered by the Virtuoso triple store~\cite{OpenLinkVirtuoso:web} available as an extension to
TNTBase~\cite{DKLRZ:PubMathLectNotLinkedData10}. In a larger Software Engineering scenario
(like a document collection of a company with multiple departments) a combination with a
Linked Data crawler, as offered by the Sponger extension to Virtuoso in an integrated server-side fashion, may have advantages: if all these departments publish their document
collections as Linked Data in the company intranet (see for
instance~\cite{Servant:LinkingEnterpriseData08} for the topicality of this example),
crawling these may reveal previously unknown connections, e.\,g.\ colleagues dealing with
structurally similar problems who could lend advice.  Note that local vocabularies
resulting from ad-hoc semantification need not be a barrier to knowledge exchange: Linked
Data practices recommend connecting occurrences of semantically equivalent resources in
different data sets by \textit{owl:sameAs}.  Alternatively, if it turns out that one
department uses a ``better'' vocabulary for their data, the \stexplus metadata extensions make
it easy to adopt it: all we have to do is to change the \stexplus bindings or |\keydef|s.\footnote{Reuse of vocabularies is not limited by traditional restrictions of {\TeX}, which has a single global namespace for macros, and where no two keys passed to a command or environment may have the same name.  {\stex} groups symbols into modules; {\stexplus} does the same for keys.  When two symbols or keys that have the same local name relatively to their module are imported into another module $M$, there are facilities for giving them distinct names for usage inside $M$.  For example, when there is already a key \texttt{name}, but the \texttt{name} property from the FOAF ontology should also be reused, we can set up a \emph{qualified import} of the latter, e.\,g.\ as \texttt{FOAFname}.}

\section{Related Work}\label{sec:related}

We have presented {\stexplus} as an extension of the {\LaTeX} language for both authoring Linked Data vocabularies and annotating semantic documents with them.  Thus, it is obviously related to other semantic extensions of {\LaTeX}.  But, when considering {\stexplus} as a text- and macro-based frontend to OWL and RDFa, it can also be compared to other ontology/vocabulary authoring and document annotation frontends, including such with graphical user interfaces.

SALT~\cite{Groza:SALT07} also allows for annotating semantic relations in
{\LaTeX} documents and exporting them as Linked Data.  SALT is restricted to a fixed set of rhetorical and bibliographical relations, plus the metadata fields of widely used document classes like LNCS, both of which it embeds as RDF annotations in the generated PDF, whereas \stexplus allows for (re)using arbitrary relations plus defining custom ones.  The target format of {\stexplus} is RDFa inside the generated \omdoc and
XHTML+MathML.  We have concentrated on that target, since it supports dynamic
interactions via our JOBAD system.  An export of the metadata relations to XMP annotations embedded in PDF should be
possible with the technology employed in SALT; we leave this to future work.

SOBOLEO~\cite{BZ:SOBOLEO10} is a lightweight graphical user interface for creating and editing
vocabularies/ontologies in OWL based on Web 2.0 tagging
approaches. In~\cite{BraunEtAl:OntologyMaturingProcessModel}, the authors evaluate its usage along their {\myCitation{Ontology Maturing Process Model}}, in which they confirm the
succeeding phases ``emergence of ideas'', ``consolidation in communities'',
``formalization'', and ``axiomatization'' in an ontology engineering process. Our observed
phases of spotting, relating and chunking essentially correspond, as the ``emergence of
ideas'' period did not apply (the documents were already created). Interestingly, the
``conso\-li\-da\-tion in communities'' phase does not only have to be thought of as a
development time: We found it reified in {\sd} like the {\vmodel} relations.
loomp is an example of a WYSIWYG editor for annotating HTML documents with terms from
vocabularies, yielding RDFa~\cite{HeeseEtAl:OneClickAnnot09}.  GUI tools traditionally
separate the task of vocabulary creation from document annotation; this also holds for SOBOLEO (responsible for the former task) and loomp (responsible for the latter).  {\stexplus}, on
the other hand, gives access to both tasks via the same interface: {\TeX} macros, which
are once declared, and once used -- possibly even in the same source file.

\section{Conclusion and Future Work}\label{sec:conclusion}

We reported on a formalization case study, where we use the \stex format, a document
formatting system and specification platform for semantic, mathematical vocabularies, on a
document corpus from Software Engineering. To cope with the the multi-dimensional
semantic structure implicit in the document collection, we extended \stex into a markup
platform for semi-formal ontologies and Linked Data called \stexplus (in our case semi-formal documents
with RDFa-based metadata annotations). 

The key observation from our case study is that if we use \stexplus as a human- and
document-oriented frontend for Linked Data documents, we can approach the formalization of
semi-formal document collections as a process of ``\emph{document and ontology
  co-development}'', where (in our case pre-existing) documents are semantically
preloaded with inter- and intra-document relations, whose meaning is given by
(project-specific or general, reusable) metadata ontologies. As we have seen in
section~\ref{sec:samsdocs}, preloading documents and developing metadata ontologies in a
joint frontend format reduces formalization barriers.  For instance, we often have to
elaborate informal document fragments into metadata vocabularies; see the discussion
about the ``rd.\@'' document state. 

For practical applicability of the \stexplus approach, machine support for
authoring and
managing \stex document collections is crucial. As a client-side counterpart to the
integrated repository and Linked Data publishing solution provided by
TNTBase~\cite{DKLRZ:PubMathLectNotLinkedData10}, we are currently developing an integrated
collection authoring environment \stexide for \stex on the basis of the Eclipse
framework~\cite{JucKoh:sidesc10}. We expect that extending \stexide to operationalize the
\stexplus functionality presented in this paper will turn it into an IDE for document
collection and ontology co-development that will enable authors to cope with the
complexities of dealing with large collections of semi-formalized documents. On the other
hand, we expect the modular \stexide system to be a good basis for deploying supportive
services in a flexible document collection environment.

We conjecture that the \stexplus based workflow for document and ontology co-development can be
extended to arbitrary Linked Data applications.

\paragraph{Acknowledgments}
The authors gratefully acknowledge the careful work of Christoph L\"uth, Holger T\"aubig,
and Dennis Walter that went into preparing the SAMS document collection, which is the
basis of this paper. Moreover, we like to thank the members of the FormalSafe project for
valuable discussions. 

\bibliographystyle{abbrv}
\bibliography{\macpathkbibs{kwarc}}
%
%
\balancecolumns
\end{document}